\documentstyle[12pt,titlepage]{article} 
\title{Exclusive Decays $B \rightarrow K \psi , B \rightarrow K^* \psi $ using
Heavy Quark Symmetries}
\author{Mohammad R. Ahmady and Dongsheng Liu}          
\date{November, 1992}   
\def\_{\rule{.3em}{.15ex}}  

\begin{titlepage}
 \begin{center}
  \vspace{0.75in}
  {\bf {\LARGE Exclusive Decays $B \rightarrow K \psi , B \rightarrow K^* \psi
$ using Heavy Quark Symmetries} \\
  \vspace{0.75in}
  Mohammad R. Ahmady and Dongsheng Liu} \\
  International Centre For Theoretical Physics \\
  P.O. Box 586 \\
  34100 Trieste, Italy\\
  \vspace{1in}
  ABSTRACT \\
  \vspace{0.5in}
  \end{center}
  \begin{quotation}
  \noindent We use the new symmetries in the heavy quark limit to calculate the
exclusive B-decays $B \rightarrow K \psi ({\psi}^{\prime})$ and $B \rightarrow
K^* \psi ({\psi}^{\prime})$.  The estimated decay rates are compared with
experimental values.
We also estimate rare $B \rightarrow K^* \gamma $ decay rate, based on these
experimental data and heavy quark symmetry.

  \end{quotation}
\end{titlepage}

\begin{document}           

\newcommand{\da}{\mbox{$\scriptscriptstyle \dag$}}
\newcommand{\lag}{\mbox{$\cal L$}}
\newcommand{\tr}{\mbox{\rm Tr\space}}
\newcommand{\fc}{\mbox{${\widetilde F}_\pi ^2$}}
\newcommand{\ns}{\textstyle}
\newcommand{\si}{\scriptstyle}

\section {Introduction}

One of the major sources of theoretical uncertainty in calculating exclusive
decays of hadrons is due to long-distance QCD which affects calculations of
hadronic matrix elements.  However, recently there has been some progress
toward model independent normalization of these matrix elements, when initial
and final state hadrons contain a heavy quark \cite{HQET,IW}.  For example, to
the leading order in $ \Lambda _{QCD} / M_Q$ ($ \Lambda _{QCD} \approx 300 MeV$
is the QCD scale and $M _Q$ is the heavy quark mass) all six form factors
parametrizing the hadronic matrix elements corresponding to $0^- \rightarrow
0^- , 1^-$ meson decay can be expressed in terms of one universal function
\cite{IW}.  This so-called Isgur-Wise function is related to the underlying QCD
dynamics and is normalized to unity when the final state meson is at rest in
the initial meson rest frame.

Whether or not this method can be applied to exclusive decays of B-meson to $ K
$ and $ K^*$ is not certain, because strange quark is not very heavy compare to
$\Lambda _{QCD}$.  But, it is tempting to apply these new symmetries in the
heavy quark limit to obtain an absolute decay rate.  In fact, in refernce
\cite{ALI}, this method has been applied to some exclusive rare B-decays.
Unfortunately, so far these decays have not been  measured accurately enough,
therefore, the extent of validity of this approach is not known.

On the other hand, exclusive B-decays to $K \psi $ and $K^* \psi $ have been
measured \cite{D}.  In this paper, we use the method of Isgur and Wise to find
the ratios of $ \Gamma (B \rightarrow K \psi ) $ and $\Gamma (B \rightarrow K^*
\psi )$ to the inclusive decay $\Gamma (B \rightarrow X_s \psi )$.  We compare
our results with the recent experimental values of these branching ratios.
Furthermore, we obtain an estimate for rare $B \rightarrow K^* \gamma$
branching ratio by the best fit of the Isgur-Wise function to these
experimental data.

\section {Exclusive decay rates $ \Gamma (B \rightarrow K \psi ) $ and $ \Gamma
(B \rightarrow K^* \psi )$ in the heavy quark limit}

We start with the effective hamiltonian relevant to the process $ B \rightarrow
\psi X_s $, $\psi$ is directly produced and $X_s$ is a hadron containing
strange quark \cite{DTP}:
\begin{equation}
 H _{eff} = C  f_{\psi} \bar s \gamma _{\mu} (1- \gamma _5 )b \epsilon ^{\mu}
_{\psi}
\end{equation}
 where $$ f_{\psi} \epsilon ^{\mu} _{\psi} = <0| \bar c \gamma ^{\mu} c| \psi >
$$ \\
\begin{equation}
 C= {{G _F} \over {\sqrt 2}} (c _1 + c _2 /3 ) V _{cs} ^* V _{cb}
\end{equation}
 $c _1 $ and $c _2$ are given in reference \cite{DTP}, taking into account
leading log corrections from short-distance QCD.  $f _{\psi}$ is obtained using
experimental results on $\psi \rightarrow e^+ e^- $.  The decay $ \Gamma (B
\rightarrow X_s \psi )$
which we equate to $ \Gamma (b \rightarrow s \psi )$ can be calculated from
(1):
\begin{equation}
\Gamma (b \rightarrow s \psi )  = \frac {C^2 f^2 _{\psi} }{8 \pi m_b m^2
_{\psi}} g(m _b , m _s , m _{\psi} ) [m^2 _b (m^2 _b + m^2 _{\psi}) - m^2 _s (2
m^2 _b - m^2 _{\psi} ) + m^4 _s -2 m^4 _{\psi}]
\end{equation}
where $$ g(m _b , m _s , m _{\psi} ) =  {\left [ {\left (1- \frac {m^2 _{\psi}
}{m^2 _b} -\frac {m^2 _s}{m^2 _b} \right )}^2 - \frac {4 m^2 _s m^2 _{\psi}
}{m^4 _b} \right ]}^{1/2}$$
In order to calculate the exclusive decay rates $ \Gamma (B \rightarrow K \psi
) $ and $ \Gamma (B \rightarrow K^* \psi )$ we need the following matrix
elements:
\begin{equation}
<K(p^{\prime} )| \bar s \gamma _{\mu} (1- \gamma _5 )b |B(p)>
\end{equation}
\begin{equation}
<K^* (p^{\prime} , \epsilon )| \bar s \gamma _{\mu} (1- \gamma _5 )b |B(p)>
\end{equation}
Using heavy quark effective theory, i.e. assuming that bottom and strange
quark are both heavy compare to $ \Lambda _{QCD} $, we can write (4) and (5) in
terms of one universal Isgur-Wise function:
\begin{equation}
<K(v^{\prime} )| \bar s \gamma _{\mu} (1- \gamma _5 )b |B(v)> = \sqrt {m _K m_
B } \xi (v. v^{\prime} ) ( v _{\mu} + v _{\mu} ^{\prime}  )
\end{equation}
\begin{equation}
\begin{array}{l}
<K^* (v^{\prime} , \epsilon )| \bar s \gamma _{\mu} (1- \gamma _5 )b |B(v)> =
\\ \sqrt {m _{K^*} m_ B } \xi (v. v^{\prime} ) \left [ (1+v. v^{\prime} )
\epsilon _{\mu} - v _{\mu} ^{\prime} ( \epsilon . v ) + i \epsilon _{\mu \alpha
\beta \gamma } v^{\alph
a} v^{\prime \beta} \epsilon ^{\gamma} \right ]
\end{array}
\end{equation}
where $ v $ and $ v ^{\prime}$ are velocities of initial and final state mesons
respectively.
 This leads to the following decay rate for $B \rightarrow K \psi $ :
\begin{equation}
\Gamma (B \rightarrow K \psi ) = \frac {C^2 f^2 _{\psi}}{16 \pi } {\left [g(m
_B , m _K , m _{\psi} ) \right ]}^3 \frac {m^2 _B {\left (m _B +m _K \right
)}^2 }{4 m _K m^2 _ {\psi} } {\vert \xi (v. v ^{\prime} ) \vert }^2
\end{equation}
where the Isgur-Wise function is evaluated at
\begin{equation}
v. v ^{\prime} = \frac {m^2 _B + m^2 _K - m^2 _{\psi} }{2 m _B m _K } \approx
3.55
\end{equation}
We use the following numerical values (in GeV) for mass parameters in our
calculations:$$
\begin{array}{lll}
m_b = 4.9        &   m_s = 0.15       \\
m_K = 0.494      &   m_{K^*} = 0.892    \\
m_{\psi} = 3.1   &   m_B = 5.28
\end{array}
$$
In the same way the $B \rightarrow K^* \psi $ decay rate can be calculated
\begin{equation}
\Gamma (B \rightarrow K^* \psi ) = \frac {C^2 f^2 _{\psi}}{16 \pi } g(m _B , m
_{K^*} , m _{\psi} )  {\vert \xi (x ) \vert }^2 m _{K^*} F(x)
\end{equation}
where
\begin{equation}
\begin{array}{l}
F(x) = (x+1) \left [ {(a+b)}^2 (x-1) +2(2x+1) \right ]
\end{array}
\end{equation}
and $$ x = \frac {m^2 _B + m^2 _{K^*} - m^2 _{\psi} }{2 m _B m _{K^*} }
\approx 2.02$$
$$ a = \frac {m_B}{m_{\psi}}, b = \frac {m_{K^*}}{m_{\psi}} $$
In the limit where b,c and s quarks are all considered heavy, the matrix
elements relevant to semileptonic $D \rightarrow K (K^*) \ell {\nu}_{\ell} $
decays can be written in terms of the same Isgur-Wise function.  For the
theoretical calculation of these decays, certain parametrizations of this
function must be adopted.  There are several different parametrizations of the
Isgur-Wise function in the literature, from which we shall adopt the monopole
and exponential forms as examples \cite{ALI}.
\begin{equation}
 \xi ( \omega ) = \frac {\omega ^2 _{\circ}}{ \omega ^2 _{\circ} -2 +2 \omega }
\end{equation}
\begin{equation}
 \xi ( \omega ) = \exp \left [ \beta (1- \omega ) \right ]
\end{equation}
where $\omega = v. v^{\prime}$ and parameters $  \omega _{\circ} \approx 1.8$
and $\beta \approx 0.5$ obtained from the best fit to data on $D \rightarrow K
\ell {\nu}_{\ell} $.  As it is pointed out in reference \cite{ALI}, (12) and
(13) lead to decay rates for $D \rightarrow K^*  \ell {\nu}_{\ell} $ which are
larger than experimental values by a factor 2. This is an indication of large $
\frac {1}{m_s}$ corrections due to not so heavy strange quark.  The range of
$\omega$ for $D \rightarrow K  \ell {\nu}_{\ell} $
decay is $\omega \in \left [ 1, \omega _{max} \approx 2.01 \right ] $, however,
for $B \rightarrow K \psi $ decay, $\xi (\omega )$ must be evaluated at the
point $ \omega \approx 3.55$ which is well beyond kinematic range of the
semileptonic D-meson decay.
  We shall continue to use (12) and (13) to make an estimate for $ B
\rightarrow K \psi $ decay.  This problem does not arise in $B \rightarrow K^*
\psi $ decay, as $x \approx 2.02$ in formula (10).  Using (12) (and (13)) we
find for the ratios:
\begin{equation}
\frac {\Gamma (B \rightarrow K \psi )}{\Gamma (b \rightarrow s \psi )} =
0.11(0.06)
\end{equation}
\begin{equation}
\frac {\Gamma (B \rightarrow K^* \psi )}{\Gamma (b \rightarrow s \psi )} =
0.54(0.53)
\end{equation}
The branching ratio for these two channels can be obtained using \cite{DTP}:
\begin{equation}
\frac {\Gamma (b \rightarrow s \psi )}{\Gamma (b \rightarrow all )} = (1.0 \pm
0.24) \times {10}^{-2}
\end{equation}

This method can also be extended to $ B \rightarrow K {\psi}^{\prime} $ and $B
\rightarrow K^* {\psi}^{\prime} $ from which the latter has been measured.  In
fact, for $B \rightarrow K^* {\psi}^{\prime} $ $ v. v^{\prime} \approx 1.62$
which is well within the validity range of (12) and (13).  Replacing $m_{\psi}$
with $m_{ {\psi}^{\prime}}$ in (3), (8) and (10) we obtain:
\begin{equation}
\frac {\Gamma (B \rightarrow K {\psi}^{\prime} )}{\Gamma (b \rightarrow s
{\psi}^{\prime} )} = 0.13(0.09)
\end{equation}
\begin{equation}
\frac {\Gamma (B \rightarrow K^* {\psi}^{\prime} )}{\Gamma (b \rightarrow s
{\psi}^{\prime} )} = 0.75(0.78)
\end{equation}
where we use $m_{{\psi}^{\prime}} = 3.69 $ GeV.  (18) together with the
theoretical ratio \cite{DTP}
\begin{equation}
\frac {\Gamma (b \rightarrow s {\psi}^{\prime} )}{\Gamma (b \rightarrow s \psi
)} = 0.32
\end{equation}
and (16) result in branching ratios $BR (B \rightarrow K {\psi}^{\prime} ) =
0.40 (0.29) \times {10}^{-3}$ and $BR (B \rightarrow K ^*{\psi}^{\prime} ) =
2.4 (2.5) \times {10}^{-3}$.  A comparison between our results and experiment
is given in table 1.\\
\\

\hskip -1cm
{\small
\begin{tabular}{c|c|c|c|c}
\hline
\multicolumn{1}{|c|}{$\begin{array}{c} Decay Mode  \end{array}$} &
\multicolumn{1}{c|}{$\begin{array}{c} Exponential Form \\ \beta = 0.5
\end{array}$} &
\multicolumn{1}{c|}{$\begin{array}{c} Monopole Form \\ \omega _{\circ} = 1.8
\end{array}$} &
\multicolumn{1}{c|}{Experiment,$B^{\pm}$} &
\multicolumn{1}{c|}{Experiment, $B^{\circ}$} \cr
\hline
\multicolumn{1}{|c|}{ $B \rightarrow K \psi $} &
\multicolumn{1}{c|}{0.06} &
\multicolumn{1}{c|}{0.11} &
\multicolumn{1}{c|}{$0.077 \pm 0.02$} &
\multicolumn{1}{c|}{$0.065 \pm 0.031$} \cr
\hline
\multicolumn{1}{|c|}{ $B \rightarrow K^* \psi $} &
\multicolumn{1}{c|}{0.53} &
\multicolumn{1}{c|}{0.54} &
\multicolumn{1}{c|}{$0.14 \pm 0.07$} &
\multicolumn{1}{c|}{$0.13 \pm 0.04$} \cr
\hline
\multicolumn{1}{|c|}{ $B \rightarrow K {\psi}^{\prime} $} &
\multicolumn{1}{c|}{0.029} &
\multicolumn{1}{c|}{0.040} &
\multicolumn{1}{c|}{ $ < 0.2 $} &
\multicolumn{1}{c|}{ $ < 0.15 $} \cr
\hline
\multicolumn{1}{|c|}{ $B \rightarrow K^* {\psi}^{\prime} $} &
\multicolumn{1}{c|}{0.25} &
\multicolumn{1}{c|}{0.24} &
\multicolumn{1}{c|}{$ < 0.35 $} &
\multicolumn{1}{c|}{$ 0.14 \pm 0.09 $} \cr
\hline
\end{tabular}
\hskip 1cm
\center{Table1. Comparison between theory and experiment for branching ratios
(in \%). }}\\

\section {Estimate of $B \rightarrow K^* \gamma $ rare B-decay}

In reference \cite{ALI}, estimates of rare B-decays are obtained using (12) and
(13).  As  mentioned by the authors, there is an uncertainty in these results
due to extrapolation of the Isgur-Wise function beyond the kinematic range it
has been parametrized for.  One way to solve this problem, is to utilize the
experimental results for the exclusive B-decays $B \rightarrow K \psi$ , $B
\rightarrow K^* \psi $ and $B \rightarrow K^* {\psi}^{\prime}$ to fix the
parameters $\omega _{\circ}$ and $\beta$ in (12) and (13).  Thus, we obtain
functions $\xi ( \omega ) $ which are valid for a wide range of $ \omega \in
\left [ 1.6,3.55 \right ] $.  In fact, for the exclusive rare B-decay $B
\rightarrow K^* \gamma $ , $ \xi ( \omega )$ has to be evaluated at $ \omega
\approx 3.04 $ , which falls inside this range.

Using the recent experimental values for the above decays, we obtain:
$$
\omega _{\circ} \approx 1.37
\begin{array}{ll}
+0.17  & \\
-0.21  &
\end{array} \\
\beta \approx 0.57
\begin{array}{ll}
+0.13  & \\
-0.12  &
\end{array}
$$
by the best fit of (12) and (13).  Consequently, according to equation (31) in
reference \cite{ALI}:
\begin{equation}
R(B \rightarrow K^* \gamma ) = \frac {m_{K^*}}{4 m_B} {\vert \xi (v .
v^{\prime} ) \vert }^2 {\left ( 1 + \frac {m_B}{m_{K^*} \right)}^2
\end{equation}
where
\begin{equation}
R(B \rightarrow K^* \gamma ) = \frac { \Gamma (B \rightarrow K^* \gamma )}{
\Gamma (B \rightarrow X_s \gamma )}
\end{equation}
we reach the following prediction
\begin{equation}
R(B \rightarrow K^* \gamma ) = 0.2
\begin{array}{ll}
+0.7  & \\
-0.8  &
\end{array}
\end{equation}
We notice that these results are independent of the specific form of the
Isgur-Wise function.  This is the consequence of fitting the fuction on a
range comprising the kinematic point of the rare $B \rightarrow K^*
\gamma$ decay.

Our estimate (22) is in agreement with that of reference \cite{AC}, i.e.
$R = (10-18) \%$ obtained from an effective hamiltonian approach based on
QCD and experimentally constrained wave function model for B-meson.  It is
also close to results based on QCD sum rules supplemented by a monopole
model \cite{DPR}.  However, R is smaller than the corresponding prediction
obtained by fitting a monopole form of the Isgur-Wise function to $D
\rightarrow K e \nu$ decay \cite{ALI}.

\section{Discussion}

In this paper, we calculated some of the exclusive nonleptonic decays of
B-meson to $K$ and $K^*$, using heavy quark symmetries.  We observe that the
theoretical prediction for $B \rightarrow K \psi$ decay is roughly in agreement
with experimantal results.  This, however, is not the case for $B \rightarrow
K^* \psi ({\psi}^{\prime})$, where theoretical predictions are almost 2-3 times
as large as the experimental values.  This discrepancy was somehow expected,
because the Isgur-Wise function we used, had been obtained using the data on
semileptonic $D \rightarrow K \ell {\nu}_{\ell}$.  The decay rate obtained for
$D \rightarrow   K^* \ell {\nu}_{\ell}$ from this function is also larger than
the experimental values by a factor 2.  These are all pointing to the fact that
s-quark is far from the symmetry limit and $ \frac {1}{m_s}$ ought to be taken
into account.

On the other hand, as we pointed out, there is an advantage in fitting a
suitably parametrized Isgur-Wise function to these measured nonleptonic
B-decays.  Particularly, this may enhance the prediction for $B \rightarrow K
\gamma$ rare decay, where $\xi (v.v^{\prime}) $ has to be evaluated at
somewhere far away from the symmetry point.  Also we are not expecting a very
large deviation from our prediction, as we obtained $\xi (\omega )$ by the best
fit to decays involving $K$ and $K^*$ in the final state.

At this preliminary stage, we can not expect much more accurate outcome.
However, our results indicate that the heavy quark symmetry does work as
the first approximation to calculate the exclusive $B \rightarrow X_s$
decays.

\vspace{0.2in}
{\bf Acknowledgements:} The authors thank Tao Zhijian for useful discussions.

\end{document}